\documentclass[twocolumn,amsmath,amssymb,pra,aps]{revtex4}
\usepackage{color}
\usepackage{multirow} 
\usepackage{graphicx}
\usepackage{epstopdf}
\usepackage{array}
\usepackage[normalem]{ulem}
\usepackage{soul}
\usepackage{booktabs}
\usepackage{natbib}

\usepackage{float}
\usepackage[normalem]{ulem}

\newcolumntype{x}[1]{>{\centering\hspace{0pt}}p{#1}}
\begin{document}
\title{Slow light based optical frequency shifter}

\author{Qian Li$^1$}
\email{qian.li@fysik.lth.se}				
\author{Yupan Bao$^1$}		
\author{Axel Thuresson$^2$}	
\author{Adam N. Nilsson$^1$}	
\author{Lars Rippe$^1$}	
\author{Stefan Kr\"oll$^1$}	
\affiliation{$^1$Department of physics, Lund University, P.O. Box 118, SE-22100 Lund, Sweden}
\affiliation{$^2$Theoretical Chemistry, Lund University, P.O.Box 124, SE-22100 Lund, Sweden}

\date{\today}

\begin{abstract}
We demonstrate experimentally and theoretically a controllable way of shifting the frequency of an optical pulse by using a combination of spectral hole burning, slow light effect, and linear Stark effect in a rare-earth-ion doped crystal. We claim that the solid angle of acceptance of a frequency shift structure can be close to $2\pi$, which means that the frequency shifter could work not only for optical pulses propagating in a specific spatial mode but also for randomly scattered light. As the frequency shift is controlled solely by an external electric field, it works also for weak coherent light fields, and can \textit{e.g.} be used as a frequency shifter for quantum memory devices in quantum communication. 

\end{abstract}

\maketitle

\section{Introduction}

Changing the passband of a filter changes its  frequency response, that is, it changes which frequency components from a signal that passes through the filter. Here we demonstrate that it is possible to change the frequency of an optical signal when the passband of a filter is dynamically changed while light propagates through it.

Manipulation of the frequency of light during propagation was first proposed and experimentally demonstrated by dynamically changing the refractive index of the medium for a light pulse propagating inside a photonic crystal cavity \cite{notomi2006PRA, preble2007NP, TanabePRL}. Later it was realized in a photonic crystal waveguide by the aid of slow light effects \cite{kampfrath2010PRA, Beggs2012PRL}. The physical mechanism behind those frequency conversion processes can be seen as the adiabatic tuning of an oscillator, which is basically analogous to the classical phenomenon  of sliding a finger along a guitar string after it has been plucked or tuning a cavity by moving the end mirror of the cavity. The refractive index changes were achieved at the presence of another strong optical field, and the relative frequency shift were proportional to the relative refractive index change, $\Delta\nu$/$\nu$ =$\Delta$n/n. However, in the present work a different physical mechanism for a frequency shift is demonstrated.
 
Previously a spectral filter  prepared in a rare-earth-ion doped crystal with on/off ratio of about $\sim$60 dB has been experimentally demonstrated \cite{beavan2013JOSAB}. This work showed that the passband of the filter can be dynamically changed by combining spectral hole burning with a linear Stark effect \cite{beavan2013JOSAB}. A filter prepared in such a crystal by spectral hole burning induces sharp dispersion across the spectral hole, hence the group velocity of an optical pulse propagating inside the crystal will be greatly slowed down and the pulse will be compressed spatially \cite{Turukhin2001PRL, walther2009PRA, zhang2012APL, Sabooni2013PRL}. 
Further study reveals that in contrast to the structural slow light effect \cite{kampfrath2010PRA, Beggs2012PRL, BoydJOSAB, JacobAOP}, the slow light effect caused by spectral engineering of the absorption profile originates from the reversible energy storage between the light and the absorbing ions in the medium via off - resonant interaction
\cite{shakhmuratov2005PRA, rebane2007JL, shakhmuratov2010JMO, boyd2011JOSAB, Eric1968PRL}. The energy of the input pulse is distributed between the  polarization of the ions, $U_{med}$, and the electromagnetic field, $U_{em}$ in the material, and the group velocity of the pulse , $v_g$, can be written as \cite{shakhmuratov2010JMO, Eric1968PRL}
\begin{equation}\label{eq:1}
v_g = \frac{c/n}{1 + U_{med}/U_{em}}
\end{equation}
 where n is the refractive index of the medium and c is the speed of light in vacuum. Therefore, the group velocity of the pulse decreases with the increase of the fraction of energy accumulated in the medium. In the case of extreme slow light effects, almost all energy will be temporarily stored in the medium in the form of ion polarization. The ions will de-polarize and the energy will be returned back to light field when the pulse exits the material. Hence, by manipulating the resonance frequency of the ions (using the linear Stark effect, for example) during the light propagation inside the crystal the frequency of the optical pulse exiting the crystal can be changed.  

To verify this concept, a $\sim$1 MHz band-pass filter where all the ions in the surrounding 18 MHz region have the same sign of the Stark effect (which is called frequency shifter or frequency shift filter) is prepared in a 10 mm long Pr$^{3+}$:Y$_2$SiO$_5$ crystal as shown in Fig. \ref{fig:filter_sketch}. A 1 $\mu$s long pulse with a frequency distribution matching the passband of the filter is sent into the crystal. It propagates with a reduced speed of $c/120 000$ and is spatially compressed to $2.5$ mm inside the crystal. According to equation (\ref{eq:1}), more than $99.9\%$ of the pulse energy is in the form of ion polarization when the light pulse propagates inside the crystal \cite{shakhmuratov2010JMO}. An external electric field is switched on while the entire pulse is inside the crystal, shifting the resonance frequency of the ions by an amount of $\Delta s$ and the pulse exiting the crystal will then also have its frequency shifted by $\Delta s$. The frequency shift is linearly proportional to the electric field applied.  Since the pulse is propagating inside the transmission window of the filter, the loss can be kept very low. Furthermore, by using two crystals oriented $90^\circ$ relative to each other \cite{ClausenPRL2012} and preparing the spectral structure throughout the entire crystals, the filter could work for any spatial input mode, including scattered, randomly polarized light \cite{zhang2012APL}. The solid acceptance angle of such a frequency shift filter could basically be $2\pi$, as described later.

Another interesting feature of the present frequency shift control is that there is no strong optical field involved during the frequency shift process as compared to the methods mentioned above \cite{preble2007NP, kampfrath2010PRA, Beggs2012PRL, TanabePRL}.
 The frequency shift is purely controlled by an external electric field applied onto the crystal, which means that it is suitable for being integrated into strongly background sensitive single photon applications.  Therefore it can be used as a highly efficient frequency shifter in quantum communication, for example for quantum memories, if the incoming photon does not  match the frequency of the quantum memory device. 

\section{Experiment}

\begin{figure}[htbp]
\centering
\includegraphics[width=\linewidth]{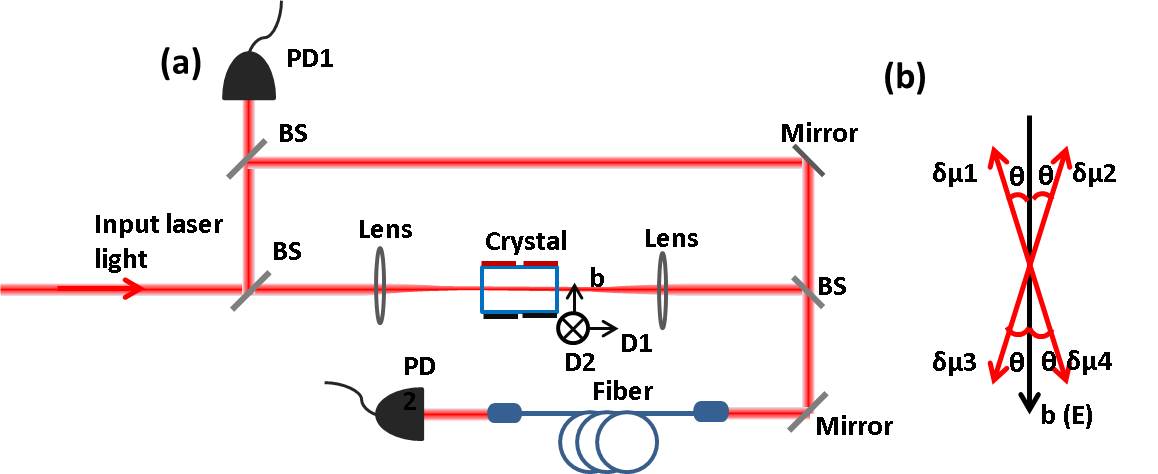}
\caption{(a). Experimental setup. Laser light polarized along the $D2$ axis of the crystal is divided into two beams by a 90/10 beam splitter (BS), where the stronger beam is focused to the center of the crystal while the weaker one is further split into two, one part is used as a reference beam to calibrate the intensity fluctuation of the laser and the other one is used as a local oscillator to beat with the delayed transmitted probe beam. Both the  transmitted beam and the local oscillator are sent into a single mode fiber for the beating signal detection. (b) The permanent electric dipole moment difference between exited state and ground state of the Pr$^{3+}$ ($\boldsymbol{\delta\mu}1,2,3,4$) relative to the electric field applied (E). The electric field is along the b axis of the crystal and the angle between the dipole moment difference and the electric field is $\theta = 12.4^{\circ}$ for all the 4 possible $\boldsymbol{\delta\mu}$.} 

\label{fig:setup}
\end{figure}

The experiment was performed in a 10 mm $\times$ 10 mm $\times$ 6 mm (crystal axes $D1 \times D2 \times b$) Pr$^{3+}$:Y$_2$SiO$_5$ crystal with $0.05\%$ doping concentration and the frequency shift filter structure was prepared  using $^3$H$_4-^1$D$_2$ transition at 2 K. This transition has an inhomogeneous linewidth of $\sim$5 GHz centered at 605.977 nm \cite{EquallPRB, Rippe2005PRA}.The lifetime of the spectral hole (filter) is about 100 seconds and can be extended to $\sim$30 minutes at the presence of a weak magnetic field ($0.01$ T) \cite{ohlsson2003PRA}. The experimental setup is shown in Fig. \ref{fig:setup} (a). A laser beam with wavelength of 605.977 nm (0 MHz in the following description) from a stabilized Coherent 699-21 ring dye laser polarized along the $D2$ axis of the crystal was split into two parts by a 90/10 beam splitter, where the stronger beam was focused onto the crystal for all the preparation and characterization of the filter and the weaker one was further split in two. One beam was sent to a photo detector ($PD1$) as a reference to calibrate the intensity fluctuations of the laser and the other one was used as a local oscillator for the heterodyne detection measurement. The light beams passing through the crystal and the local oscillator were overlapped on a beam splitter and sent through an optical fiber, and detected by a photo detector ($PD2$).

\begin{figure}[htbp]
\centering
\includegraphics[width=\linewidth]{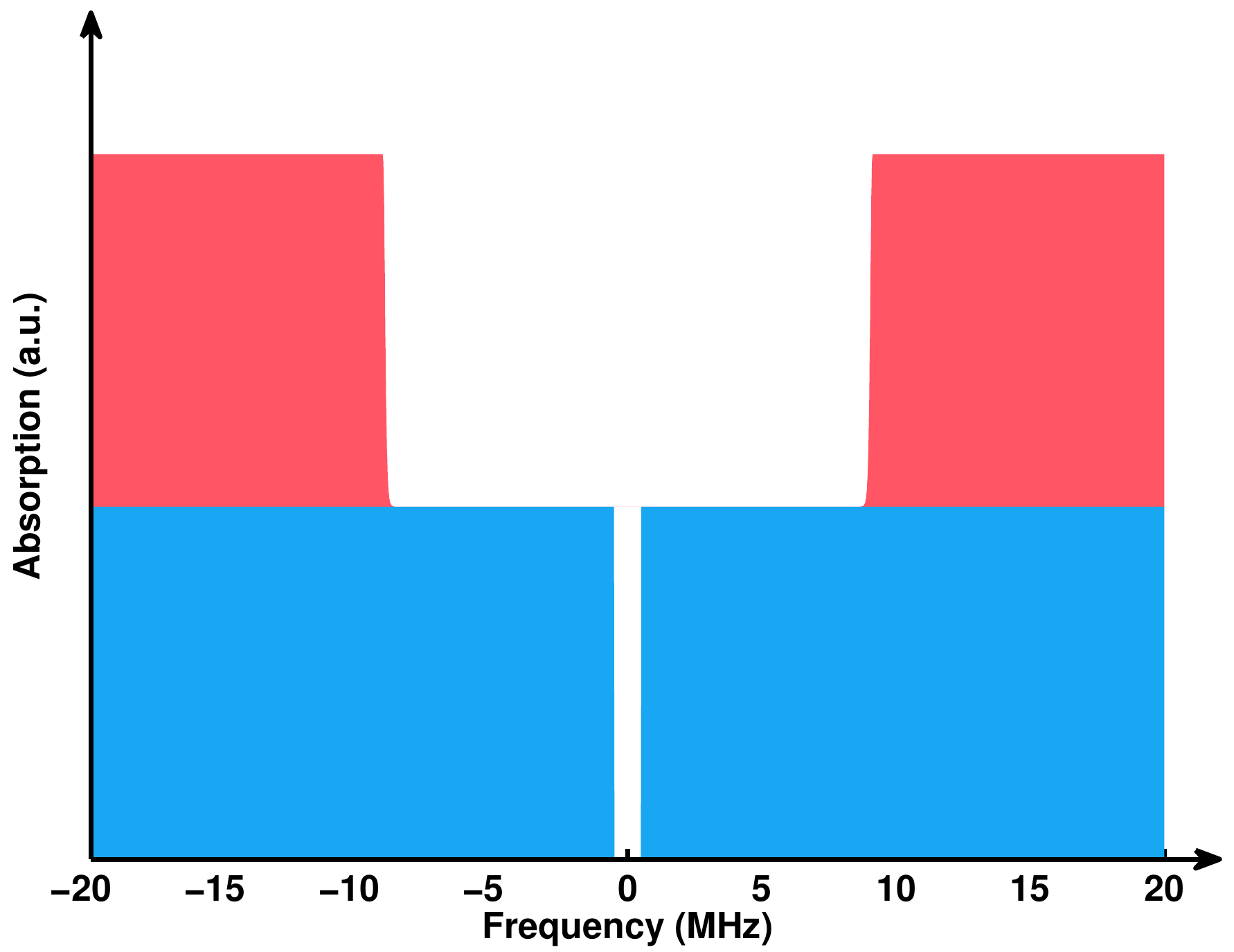}
\caption{ Frequency shift filter structure. The blue color represents ions with positive sign of the Stark effect while red color represents the ions with negative sign of the Stark coefficient.  The material has been prepared such that all ions within the $\pm$9 MHz region have the same sign of their Stark coefficient.}  

\label{fig:filter_sketch}
\end{figure}

For Pr$^{3+}$, the permanent electric dipole moment of the ground state is different from that of the excited state. Therefore applying an external electric field, \textbf{E}, will Stark shift the resonant frequency of the ions, compare \textit{e.g.} \cite{graf1997PRB}, from $\nu_0$ to $\nu'= \nu_0 + \boldsymbol{\Delta}s$, where
\begin{equation}\label{eq:2}
\boldsymbol{\Delta}s = \frac{(\boldsymbol{\mu}_e - \boldsymbol{\mu}_g) \cdot \mathbf{E}}{\hbar} = \frac{\boldsymbol{\delta\mu}  \cdot \mathbf{E}}{\hbar}
\end{equation}
where  \textbf{E} is the applied electric field, $\boldsymbol{\mu}_g$ and $\boldsymbol{\mu}_e$ are the electric dipole moments for ground state and excited state respectively, and $\boldsymbol{\delta\mu}$ is the difference between the excited and ground state dipole moments. There are four possible $\boldsymbol{\delta\mu}$ orientations for Pr$^{3+}$ as shown in Fig. \ref{fig:setup} (b), all four orientations of  $\boldsymbol{\delta\mu}$ have an angle of $\theta = 12.4^{\circ}$ relative to the b axis of the crystal, and  the magnitude of $\boldsymbol{\delta\mu}/\hbar$  is $111.6$ kHz/(V cm$^{-1}$) \cite{graf1997PRB}. The E field is applied along the b axis and  because of the symmetry of $\boldsymbol{\delta\mu}$ and E field, there will be two effective Stark coefficients for Pr$^{3+}$, with the same amplitude but opposite signs. When an external voltage is applied across the crystal, half of the ions, those with positive Stark coefficient, will shift to higher frequencies  while the rest will shift to lower frequencies.

Optical pumping together with electric field induced Stark shift were used to prepare the filter structure. In the end a $\sim$1 MHz band-pass filter surrounded only by ions with positive sign of the Start coefficient was prepared  over an 18 MHz region in the center of the inhomogeneous profile. A sketch of the final filter structure can be found in Fig. \ref{fig:filter_sketch}. More information about how this structure was created can be found in the Supplementary Material.

\section{Simulation}

A modified version of the Maxwell-Bloch (MB) equations for a 2-level system \cite{Axelthesis, foot2004atomic, cornish2000Thesis} is used to include a Stark shift where the resonance frequency of $50\%$ of the ions are shifted to higher frequencies while the rest of the ions are shifted an equal amount to lower frequencies when an electric field is applied. A simple and efficient way to simulate this effect is to introduce two Bloch vectors (denoted $\bar{r}^A$ and $\bar{r}^B$), for the two different ion groups with opposite signs of their Stark coefficient, at each frequency and position, instead of the traditional single Bloch vector. At a particular time, $\tau_0$, the resonance frequency for $\bar{r}^A$ and $\bar{r}^B$ is shifted by $+\Delta_s$ and $-\Delta_s$ respectively. The modified MB equations are then expressed as:
\begin{eqnarray}
&& \frac{\mathrm{d} r^A_{x}}{\mathrm{d}\tau} = - (\Delta+\Delta_s\Theta(\tau-\tau_0)) r^A_{y} -\Omega_ir^A_{z}-\frac{r^A_{x}}{T_2}
\\
&& \frac{\mathrm{d} r^A_{y}}{\mathrm{d}\tau} = (\Delta+\Delta_s\Theta(\tau-\tau_0)) r^A_{x} + \Omega_rr^A_{z}-\frac{r^A_{y}}{T_2}
\\
&& \frac{\mathrm{d} r^A_{z}}{\mathrm{d}\tau} = \Omega_ir^A_{x} - \Omega_rr^A_{y}-\frac{1+r^A_{z}}{T_1}
\\
&& \frac{\mathrm{d} r^B_{x}}{\mathrm{d}\tau} = - (\Delta-\Delta_s\Theta(\tau-\tau_0)) r^B_{y} -\Omega_ir^B_{z}-\frac{r^B_{x}}{T_2}
\\
&& \frac{\mathrm{d} r^B_{y}}{\mathrm{d}\tau} = (\Delta-\Delta_s\Theta(\tau-\tau_0)) r^B_{x} + \Omega_rr^B_{z}-\frac{r^B_{y}}{T_2}
\\
&& \frac{\mathrm{d} r^B_{z}}{\mathrm{d}\tau} = \Omega_ir^B_{x} - \Omega_rr^B_{y}-\frac{1+r^B_{z}}{T_1}
\\
&& \frac{\mathrm{d}\Omega_r}{\mathrm{d}z'}=\frac{\alpha_0}{2 \pi}\bigg( \int_{-\infty}^{+\infty}g^A r^A_{y}\mathrm{d}\Delta+\int_{-\infty}^{+\infty}g^B r^B_{y}\mathrm{d}\Delta \bigg)
\\
&& \frac{\mathrm{d}\Omega_i}{\mathrm{d}z'}=-\frac{\alpha_0}{2 \pi}\bigg( \int_{-\infty}^{+\infty}g^A r^A_{x}\mathrm{d}\Delta+\int_{-\infty}^{+\infty}g^B r^B_{x}\mathrm{d}\Delta \bigg)
\end{eqnarray}
where $\bar{r}^A = $($r^A_{x}$, $r^A_{y}$, $r^A_{z}$) , $\bar{r}^B = $($r^B_{x}$, $r^B_{y}$, $r^B_{z}$), $g^A$  and $g^B$ are the spectral profiles (including the transmission windows) for $\bar{r}^A$ and $\bar{r}^B$ respectively. $\Omega_r$ and $\Omega_i$ are the real and imaginary Rabi frequencies, $\Delta$ is the detuning, $\tau$ is retarded time, $z'$ is the position in the one-dimensional crystal, $\alpha_0$ is the absorption coefficient, $T_1$ and $T_2$ are the  decay constants for the life time and coherence of the excited state respectively, and $\Theta$ is the Heaviside step function which is used to describe the shift in frequency when the electric field is turned on at time $\tau_0$. Any function properly describing the frequency shift of the ions due to the change of the electric field could be used instead of the Heaviside function.

\section{Results and Discussion}

\begin{figure}[htbp]
\centering
\includegraphics[width=\linewidth]{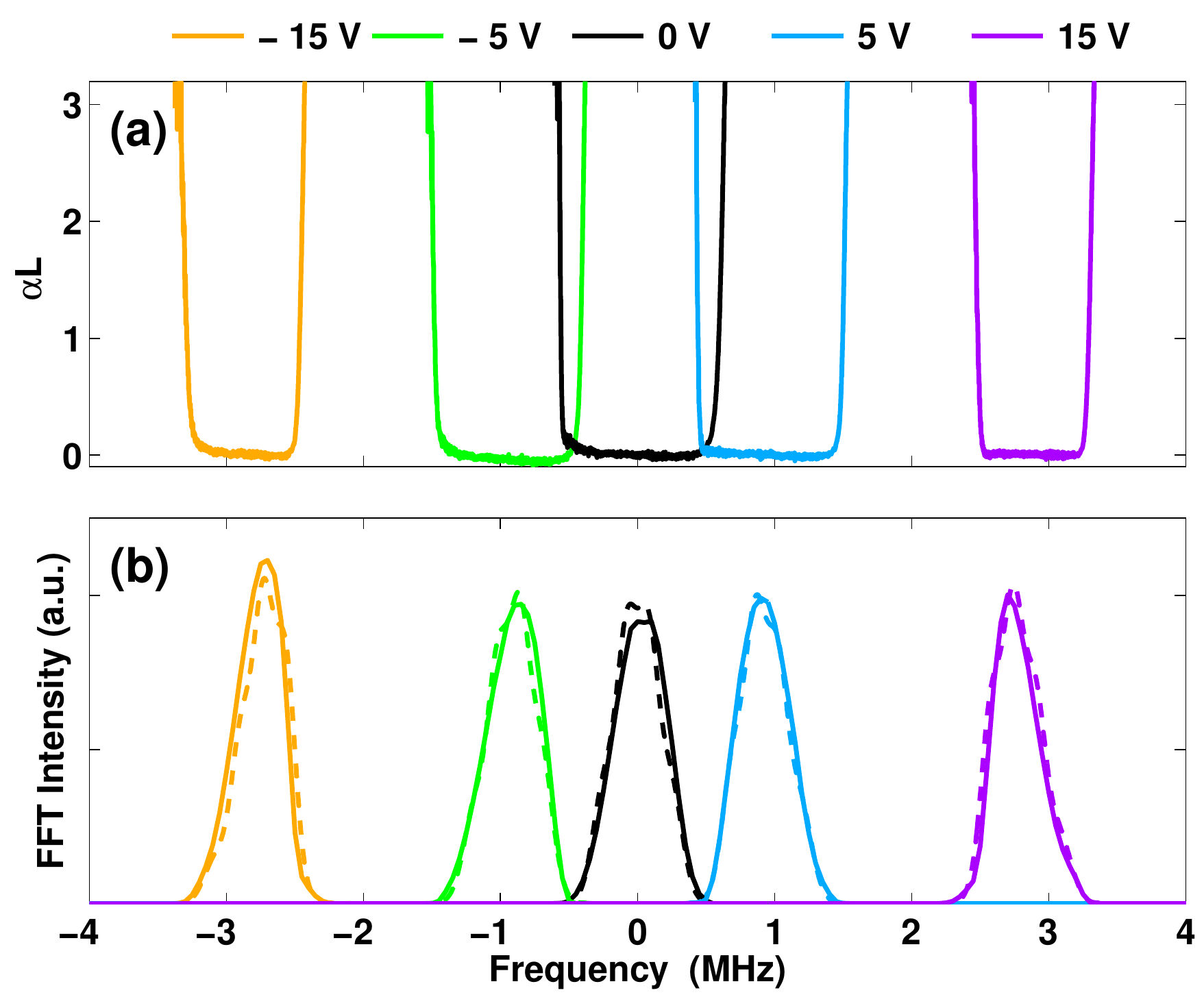}
\caption{(a) The passband position of the frequency shift filter when different voltages are applied. The black trace is for the filter created originally. When different voltages are applied across the crystal, the resonance frequency of the ions outside the filter shifts, therefore the passband of the filter changes accordingly. (b) Frequency distribution of the transmitted pulse for different voltages, measured experimentally (solid traces) and simulated theoretically (dashed traces). The incoming probe pulse is at 0 MHz (605.977 nm), and different voltages are applied when the incoming pulse is completely inside the crystal. The frequency shift of the transmitted pulse follows the shift of the filter and is proportional to the external voltage applied. }
\label{fig:Readout_FFT}
\end{figure}

After the structure preparation, a frequency chirped pulse with a chirp rate of 1 kHz/$\mu$s was sent into the crystal to map out the structure of the frequency shift filter. The intensity of this pulse was reduced until optical repumping of the ions was too small to affect the spectral structure. The black trace in Fig. \ref{fig:Readout_FFT} (a) shows the initially created filter structure. Since the ions around the filter all have a positive sign of the Stark coefficient, the passband of the filter will shift to higher frequencies when a positive external voltage is applied and vice versa as can be seen in Fig. \ref{fig:Readout_FFT} (a). The filter becomes narrower when higher voltages are applied onto the crystal, especially on the frequency side further away from zero frequency. There could be three possible reasons for this. One is the inhomogeneity of the electric field (Supplementary Material) along the pulse propagation direction, which causes the frequency shift of the ions to be slightly different. While most of the ions shift the same amount in frequency, some ions shift less. Another reason is that the ions with the opposite signs of the Stark coefficient moves closer to the filter as higher voltages are applied and since the ions have a Lorentzian absorption profile, there will be some absorption caused by these ions.  Both of these two effects will make the filter narrower with higher electric field and have larger effect on the frequency edge that is further away from the zero frequency. The third possible reason is that the E field applied is not exactly along the $b$ axis of the crystal, so that the projection of $\boldsymbol{\delta\mu}$ of the selected ion onto the E field is not the same anymore. Therefore, when a certain E field is applied, half of the ions will be shifted more than the rest of the ions. Since this misalignment has the same effect on both side of the hole structure, it seems unlikely to be the main reason here. 

Having the frequency shift filter prepared, a 1 $\mu$s long Gaussian pulse with center frequency at 0 MHz was sent into the crystal. Due to the off-resonant interaction between the optical pulse and the absorbing ions outside the filter, the group velocity of the probe pulse is slowed down to about 2500 $m/s$ and it takes about 4 $\mu$s for the pulse to pass through the crystal. According to equation (\ref{eq:1}), with this group velocity, more than $99.9\%$ of the pulse energy is stored in the form of ion polarization when the pulse is inside the crystal. An external voltage is applied to shift the resonance frequency of the ions and the frequency of the transmitted pulse will shift as a consequence. Heterodyne detection was used to measure the frequency of the transmitted pulse. The solid curves in Fig. \ref{fig:Readout_FFT} (b) shows the frequency distribution of the transmitted pulses for different voltages measured experimentally, while the dashed curves are from the theoretical simulation for the same process. The black trace is for the case where no external voltage is applied, hence no shift is observed. A Stark coefficient of 116.7 kHz/V/cm is used in the simulation, which fits well with the literature value reported before \cite{graf1997PRB, Morganthesis}. Since the frequency shift of the pulse originates from the frequency shift of the ions as the energy of the pulse is stored in the ions during the propagation inside the crystal, it is quite intuitive that the frequency shift of the pulse goes together with the frequency shift of the filter. 

\begin{figure}[htbp]
\centering
\includegraphics[width=\linewidth]{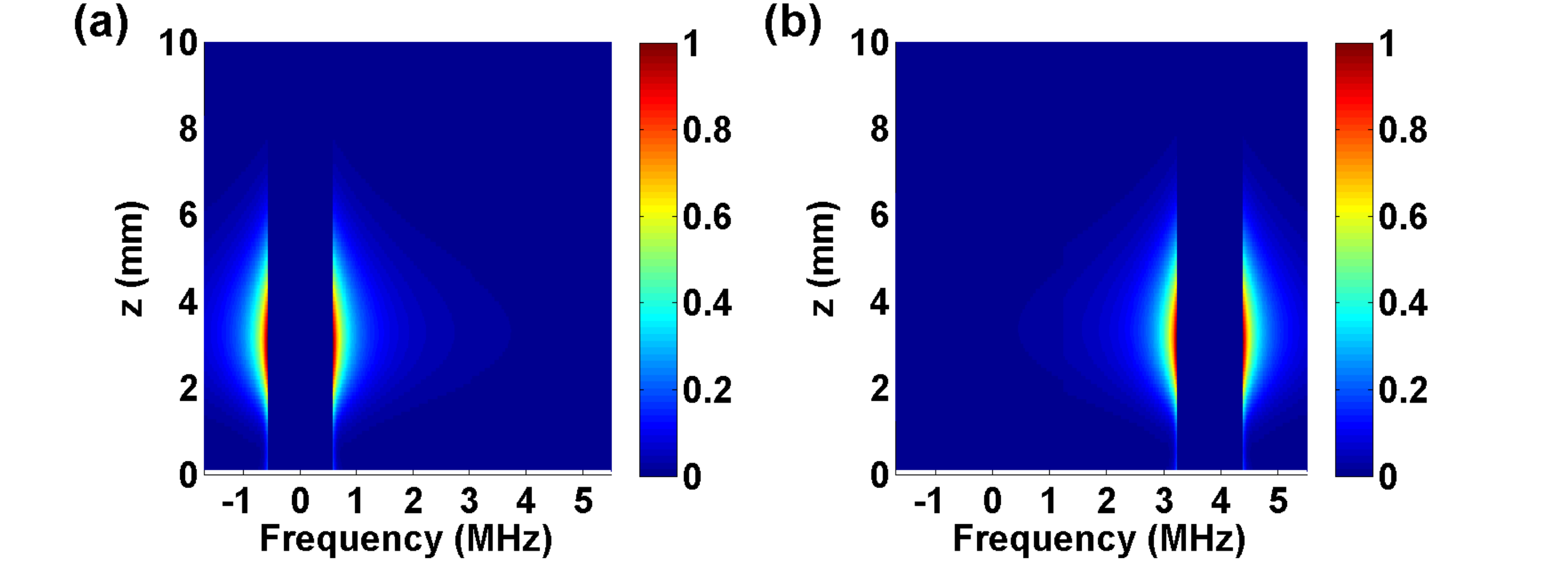}
\caption{The energy distribution in the ions just before (a) and after (b) applying electric field. The energy of the pulse is stored in the ions outside the $\sim$1 MHz transmission window, and the frequency shift of the ions will shift the light frequency. }
\label{fig:pulseenergy}
\end{figure}

To illustrate the frequency shift process inside the crystal, the energy distribution just before and after the frequency shift is shown in Fig. \ref{fig:pulseenergy}, where the dark red color represents most of energy is stored while the dark blue represents no energy is stored at that frequency and position (a full video of the energy flow in the crystal during the pulse propagation can be found in the ancillary files to this article). The energy distribution is calculated by multiplying the number of ions at a specific frequency and position by the probability of them being excited and is represented by different color. It can be seen from Fig. \ref{fig:pulseenergy} (a), the empty $\sim$1 MHz channel around 0 MHz is the transmission window originally created, and even though the input pulse itself only contains energy within the $\sim$1 MHz passband, the energy of the pulse is stored in the ions outside the passband. After applying an electric field, all the ions are shifted 3.8 MHz as shown in Fig. \ref{fig:pulseenergy} (b), which will give a light pulse at a new frequency in the end. 

\begin{figure}[htbp]
\centering
\includegraphics[width=\linewidth]{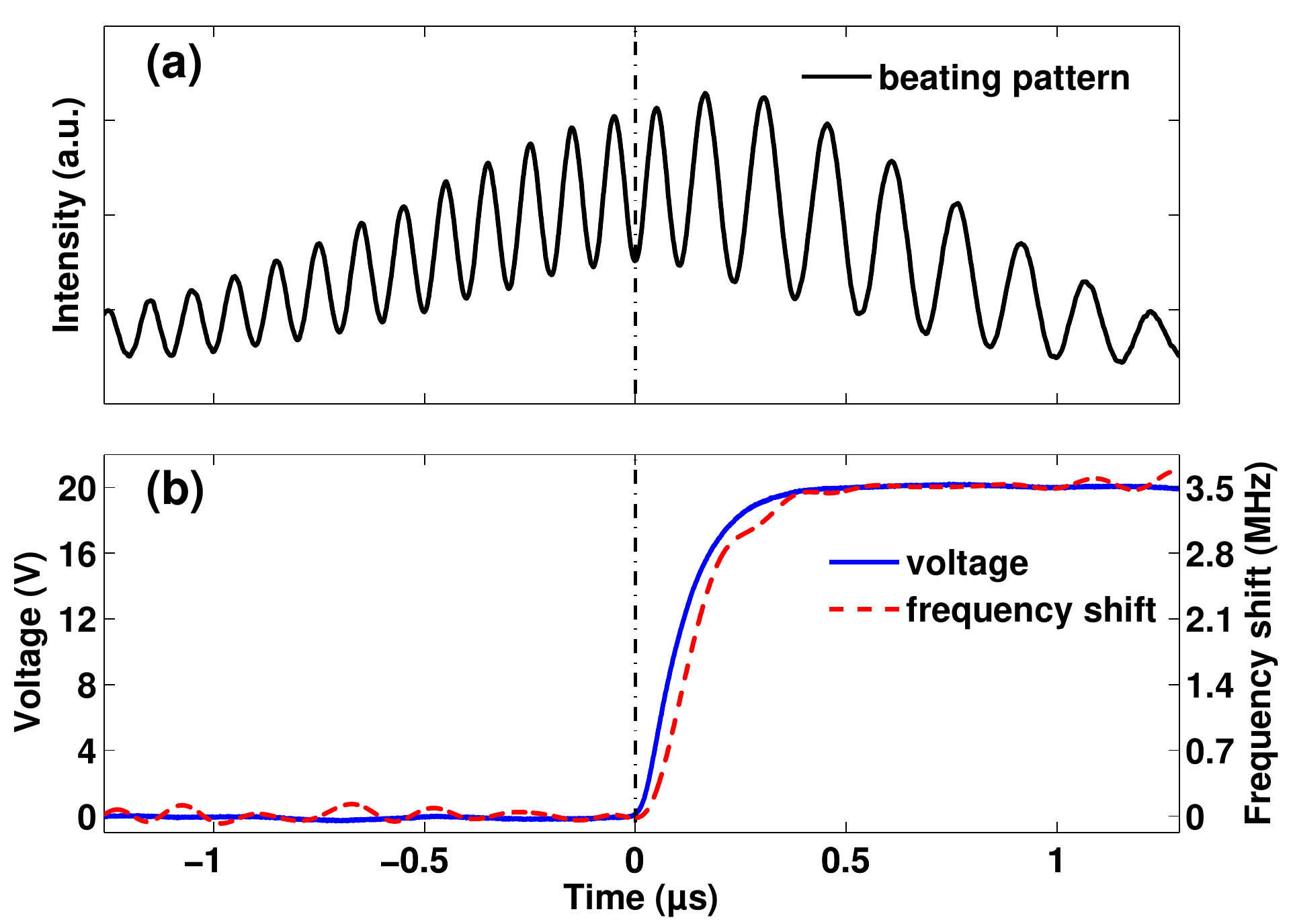}
\caption{(a). Beating pattern from the transmitted pulse after the crystal and the reference beam. The electric field is switched on at time t = 0, as indicated by the dash-dotted line. The change of beating frequency illustrates that the frequency of the transmitted pulse is changed. (b). The instantaneous frequency calculated from the beating frequency of the transmitted beam (red, dashed line) is plotted together with the external voltage (blue, solid line) applied onto the crystal, the frequency shift follows the voltage change. }
\label{fig:frequencyshift}
\end{figure}

\begin{figure}[htbp]
\centering
\includegraphics[width=\linewidth]{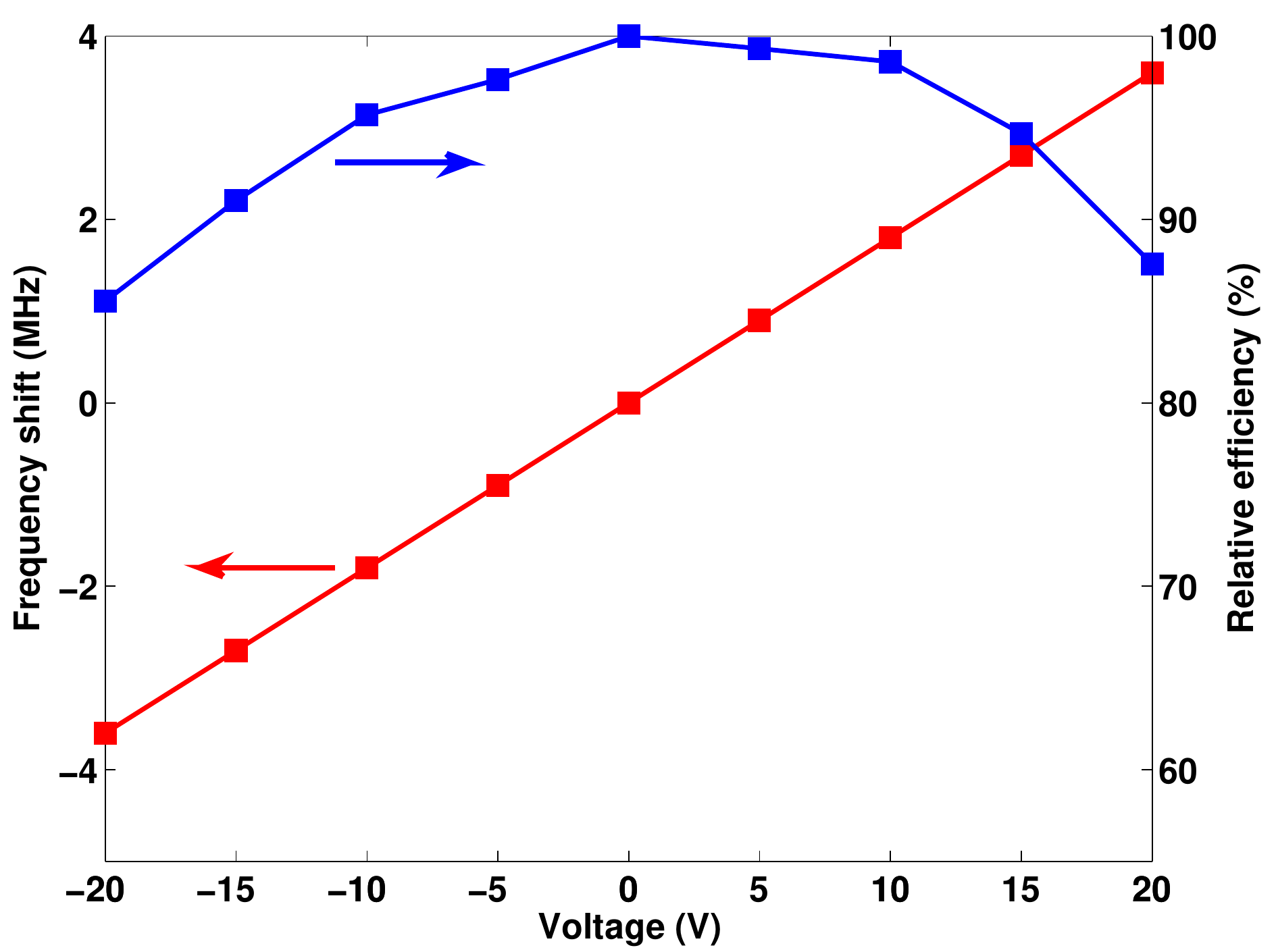}
\caption{ Frequency shift of the transmitted pulse (red) and relative efficiency of the frequency shift (blue) for different voltages applied. }
\label{fig:volt_FFT_loss}
\end{figure}

To further study the frequency shift process experimentally, an external electric field was applied when half of the optical pulse had exited the crystal while the other half is still inside. As can be seen from the beating pattern in Fig. \ref{fig:frequencyshift} (a), the phase evolution of the beating pattern changes as the electric field is switched on, the first and second half of beating pattern have different frequencies, which shows that there is a frequency shift of the transmitted pulse after the electric field is switched on. The instant frequency of the transmitted pulse is calculated as described in \cite{TakedaJOSA, RippeArxiv} and plotted together with the applied voltage as shown in Fig. \ref{fig:frequencyshift} (b). The frequency shift follows the voltage applied and the switch on time of the frequency shift filter is about 200 ns, limited by the rise time of the external voltage. 
 It can be confirmed from the red trace of Fig. \ref{fig:volt_FFT_loss} that the frequency shift of the transmitted pulse is proportional to the external voltage applied and can be well controlled electrically.

Finally, we study the efficiency of the frequency shift process. The relative efficiency, i.e. $ 
\eta_{rel} = \frac{Energy(V)}{Energy(V=0)}$ , where $Energy(V)$ is the energy of the transmitted pulse for a specific voltage (electric field). To this end, the transmitted, frequency shifted pulse from the filter was averaged 50 times and measured directly by $PD2$, the local oscillator beam was blocked (see Fig. \ref{fig:setup}). The relative efficiency was calculated by comparing the pulse area of the averaged transmitted pulse at different voltages with the case where no voltage is applied. The result is shown as the blue trace in Fig. \ref{fig:volt_FFT_loss}. For a frequency shift of less than $\pm3$ MHz (applied voltage $\pm17$ V) the relative efficiency is above $90 \%$. The larger the frequency shift, the lower the efficiency.  This can be explained by the narrowing of the filter structure after applying the electric field as shwon in Fig. \ref{fig:Readout_FFT}. The narrowing after the frequency shift induces absorption for the outer frequency components of the light and causes the decrease in efficiency.

If we assume that the main loss inside the 1 MHz filter comes from the off-resonance excitation of the ions outside the hole and that the optical depth outside the hole is 20 for this case (estimated from \cite{Rippe2005PRA, MahmoodPRL_cavity, MahmoodPolarization}), then the fundamental limit on how much absorption remains in the middle of a 1 MHz inverse top hat distribution of Pr$^{3+}$ ions, which have a homogeneous linewidth of 1 kHz, can be calculated to be $\sim$2$\%$. However, such low loss has not been experimentally demonstrated for some similar case\cite{beavan2013JOSAB}.

\begin{figure}[htbp]
\centering
\includegraphics[width=\linewidth]{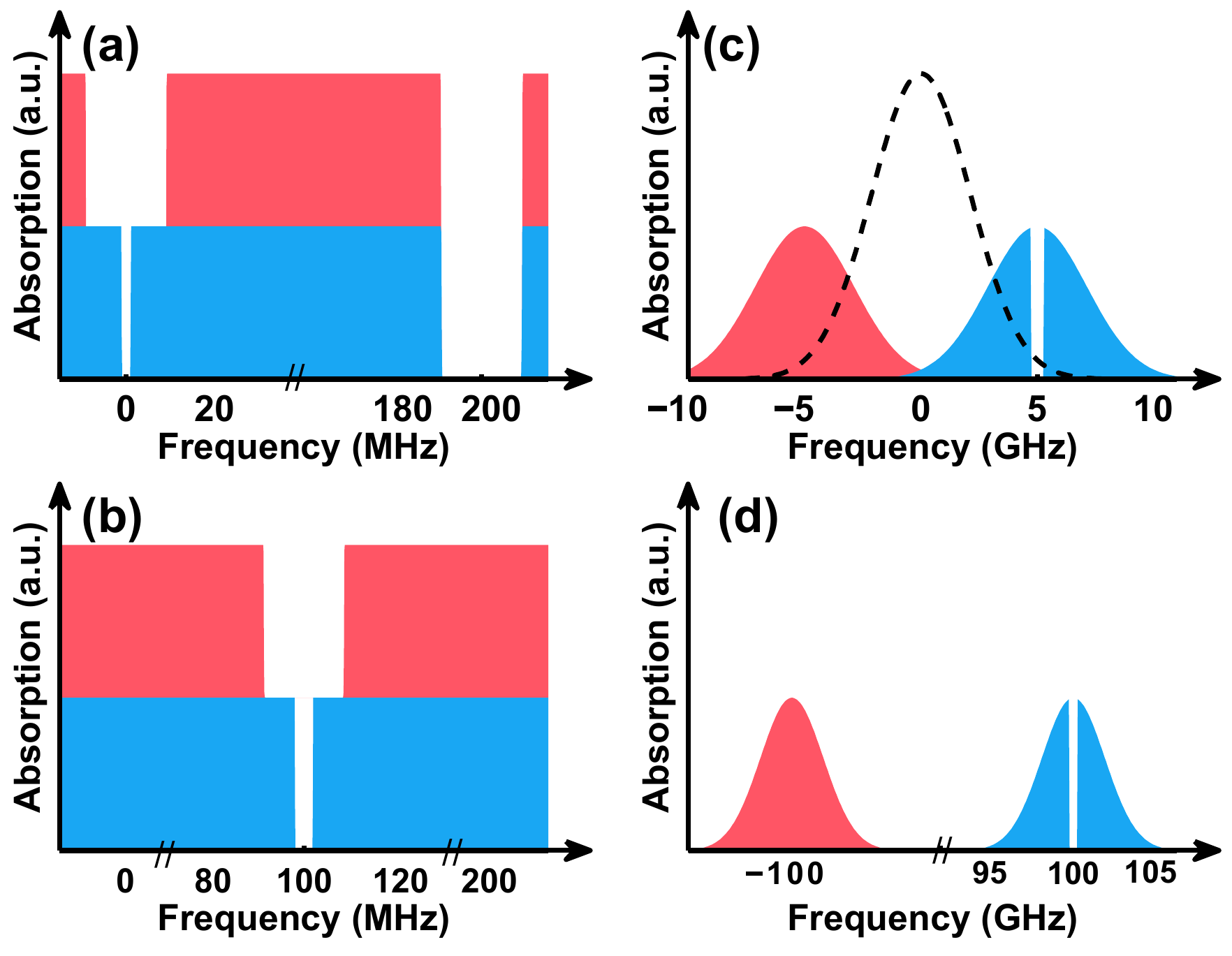}
\caption{ Ways to make larger frequency shift. The red color stands for the ions with negative Stark coefficient while the blue ones are ions with positive Stark shift. (a) A frequency shift filter structure at 0 MHz and another spectral hole at 200 MHz away. (b) After applying a certain E field, the frequency shift filter moves to 100 MHz. Using this method the frequency shift of an optical pulse could be hundreds of MHz. (c) The black dashed line is the original absorption profile with FWHM of 5 GHz. A static E field can completely separate the ions with different signs of their Stark coefficient. The frequency shift filter will be prepared by simple hole burning in one group of these ions. The width of the spectral hole is exaggerated here. (d) Applying an additional E field will shift the ions as well as light propagating inside the medium up to hundreds of GHz.}
\label{fig:100MHz_filter}
\end{figure}

Currently the frequency shift is about $\pm4$ MHz in this experiment and is limited by our choice of hole burning technique and the existence of the two groups of Pr$^{3+}$ with different signs of their Stark coefficient. If shifted further, there will be absorption from ions with opposite signs of their Stark shift. Nevertheless, one can go beyond this limit. For example, by preparing another spectral hole at twice the desired frequency shift, the frequency shift can be extended to hundreds of MHz. As shown in Fig. \ref{fig:100MHz_filter} (a), besides the frequency shift filter prepared at zero frequency, another wide hole is created at 200 MHz. Applying an E field can shift the red ions  $-100$ MHz while shifting the blue ions $+100$ MHz. In the end, the frequency shift filter at 0 MHz and the spectral hole at 200 MHz will overlap at 100 MHz as shown in Fig. \ref{fig:100MHz_filter} (b). Therefore, the optical pulse propagating inside the crystal can be shifted from 0 MHz to 100 MHz in this way. We propose that the loss during the frequency shift process can be estimated as follows. Looking at Fig. \ref{fig:100MHz_filter} (a), the "blue" ions which are polarized by the input pulse will need to shift towards higher frequency until they meet the spectral hole from the "red" ions at the 100 MHz position (Fig. \ref{fig:100MHz_filter} (b)). During the shift from zero to 100 MHz before they reach the red spectral hole, the pulse in the blue transmission window can be absorbed by the "red" ions between 10 MHz and 90 MHz (Fig. \ref{fig:100MHz_filter} (a) and (b)). The losses, during this process will be, 
\begin{equation}\label{eq:3}
\eta \approx 1- e^{-\alpha/2\times L'}
\end{equation}
where $L' = v_g \cdot T$ is the distance in the crystal that the pulse propagates during the electric field switching time, $T$. 
\begin{equation}\label{eq:4}
v_g \approx \frac{2\pi\Gamma}{\alpha/2}
\end{equation}
is the group velocity, $\Gamma$ is the width of the filter transmission window, and $\alpha/2$ is the absorption coefficient of half of the ions. In equation (\ref{eq:3}) $\alpha/2$ is the "red" ion absorption while in equation (\ref{eq:4}) $\alpha/2$ is the "blue" ion absorption. Inserting equation (\ref{eq:4}) in equation(\ref{eq:3}), we get,
\begin{equation}\label{eq:5}
\eta \approx 1 - e^{-2\pi\Gamma\cdot T}
\end{equation}
The loss is only affected by the width of the filter and the  switch on time of the voltage. For $\Gamma = 1$ MHz, a loss about $3\%$ per 5 ns switching time would be achieved. So the loss can be kept low if the field switch on is fast enough. This way the frequency shift is only limited by the frequency range of the laser used for the structure creation. 

Moreover, one can extend the frequency shift range still further by applying a large E field to completely separate the ions with opposite sign of their Stark coefficients. In this way, the absorption from the other group of ions can be completely eliminated, and will no longer affect the efficiency for the frequency shift process. For instance, for a 20 $\mu$m thick waveguide made out of Pr$^{3+}$:Y$_2$SiO$_5$, a static field of 100 V will be sufficient to separate the absorbing Pr$^{3+}$ with different signs of Stark coefficient 10 GHz apart (the inhomogeneous linewidth is about 5 GHz), as shown in Fig. \ref{fig:100MHz_filter} (c). Then by doing the straightforward hole burning in one group of ions together with the Stark effect, frequency shifts up to a couple of hundred GHz can be achieved, see Fig. \ref{fig:100MHz_filter} (d).  
In this case the frequency shift can reach hundreds of gigahertz and is only limited by the breakdown voltage of the crystal ($\sim$MV/cm). Note that, here the homogeneity of the E field along the light propagation direction has to be good enough to not smear out  the hole structure.

A band-pass filter can be prepared throughout the entire crystal if the preparation beam covers or scans the whole crystal \cite{zhang2012APL}. It has been shown that, a filter prepared in this way will work for light entering the crystal from multiple directions. The light will be transmitted and delayed if its frequency is within the passband, and strongly attenuated otherwise \cite{zhang2012APL}. By using special anti-reflection coating techniques, the input light can be coupled into the crystal very effectively even with a high incidence angle \cite{spinelliNature2012, GuoOE2014}. Once inside the crystal, the light will be guided to the exit facet via total internal reflection (the refractive index for the crystal is n $\approx 1.8$). Therefore the solid angle of acceptance of such a filter could be close to 2$\pi$. 
In the present demonstration, the input light is polarized along the $D2$ axis of the crystal since this is the axis with maximum slow light effect (energy storage). For light entering the crystal polarized perpendicular to the $D2$ axis the slow light effect would be so small that the light just passes through before there is time to change the electric field and the frequency shift can not be achieved. This polarization restriction can be eliminated by adding an identical crystal right after the present one rotated around the light propagation axis ($D1$) by  $90^\circ$ relative to the first one \cite{ClausenPRL2012}. A frequency shift filter made out of two identical crystals and the same spectral structure prepared in the entirety of both crystals should be able to shift the frequency of light in any spatial mode and any polarization, including randomly scattered light.

\section{Conclusion}

In conclusion, we demonstrate a voltage (electric field) controlled optical frequency shift device capable of shifting the frequency of a light pulse by $\sim$1 MHz/(V/mm). The device is based on creating a narrow semi-permanent transmission window in frequency in a rare-earth-ion doped crystal absorption line using optical pumping. This causes strong slow light effects (group velocities of a few thousand m/s) for light propagating in this transmission window. At these slow group velocities the energy of an propagating optical pulse is mainly stored as optical polarization of the ions in the material. By Stark shifting these ions, the frequency of the light pulse is shifted. The demonstrated frequency shifting range is $\pm$4 MHz and the fractional pulse energy loss when applying the frequency shift is in the a few to ten percent range. It is proposed how the frequency range of these shifts could be extended to be limited only by the breakdown voltage of the crystal ($\sim$1 MV/cm) yielding a frequency shift limit of around 100 GHz. Based on related work \cite{zhang2012APL, ClausenPRL2012}, it is argued that these frequency shifters should have a solid acceptance angle close to $2\pi$ and thus could be used also for scattered light. As the frequency shift is controlled solely by an external electric field and does not involve any additional optical pulses, these frequency shifters can be especially suitable in weak light situation, for example, to match the frequency of weak coherent light to a particular quantum memory frequency. Additionally, it has a potential to be integrated with other devices, which could open up on-chip applications.
 
\section{Acknowledgement}
This work was supported by the Swedish Research Council, the Knut $\&$ Alice Wallenberg Foundation. The research leading to these results also received funding from the People Programme (Marie Curie Actions) of the European Union's Seventh Framework Programme FP7 (2007-2013) under REA grant agreement no. 287252 (CIPRIS) and Lund Laser Center (LLC).

\bibliography{FrequencyShifter_arXiv}
 \clearpage
\begin{center}
\textbf{\large Supplemental Materials: Slow light based optical frequency shifter}
\end{center}
\setcounter{equation}{0}
\setcounter{figure}{0}
\setcounter{table}{0}
\setcounter{page}{1}
\makeatletter
\renewcommand{\theequation}{S\arabic{equation}}
\renewcommand{\thefigure}{S\arabic{figure}}
\renewcommand{\bibnumfmt}[1]{[S#1]}
\renewcommand{\citenumfont}[1]{S#1}

\subsection{Structure preparation}

For the frequency shifter to work, it is critical that the absorbing ions right outside the filter should have the same sign of their Stark coefficient. However, in Pr$^{3+}$:Y$_2$SiO$_5$, there are four possible static dipole orientations, distributed (in the $D2-b$ plane) $12.4^{\circ}$  away from b axis (Fig. \ref{fig:setup} (b) in the main text). For an electric field applied along the b axis, this yields two overall Stark coefficients, with the same magnitude but opposite sign. To make the frequency shifter structure Fig. \ref{fig:filter_sketch} in the main text, the ions with different signs of their Stark coefficient have to be separated. This is done by combining the hole burning technique together with the Stark effect during the structure preparation process.

\begin{figure}[htbp]
\centering
\includegraphics[width=\linewidth]{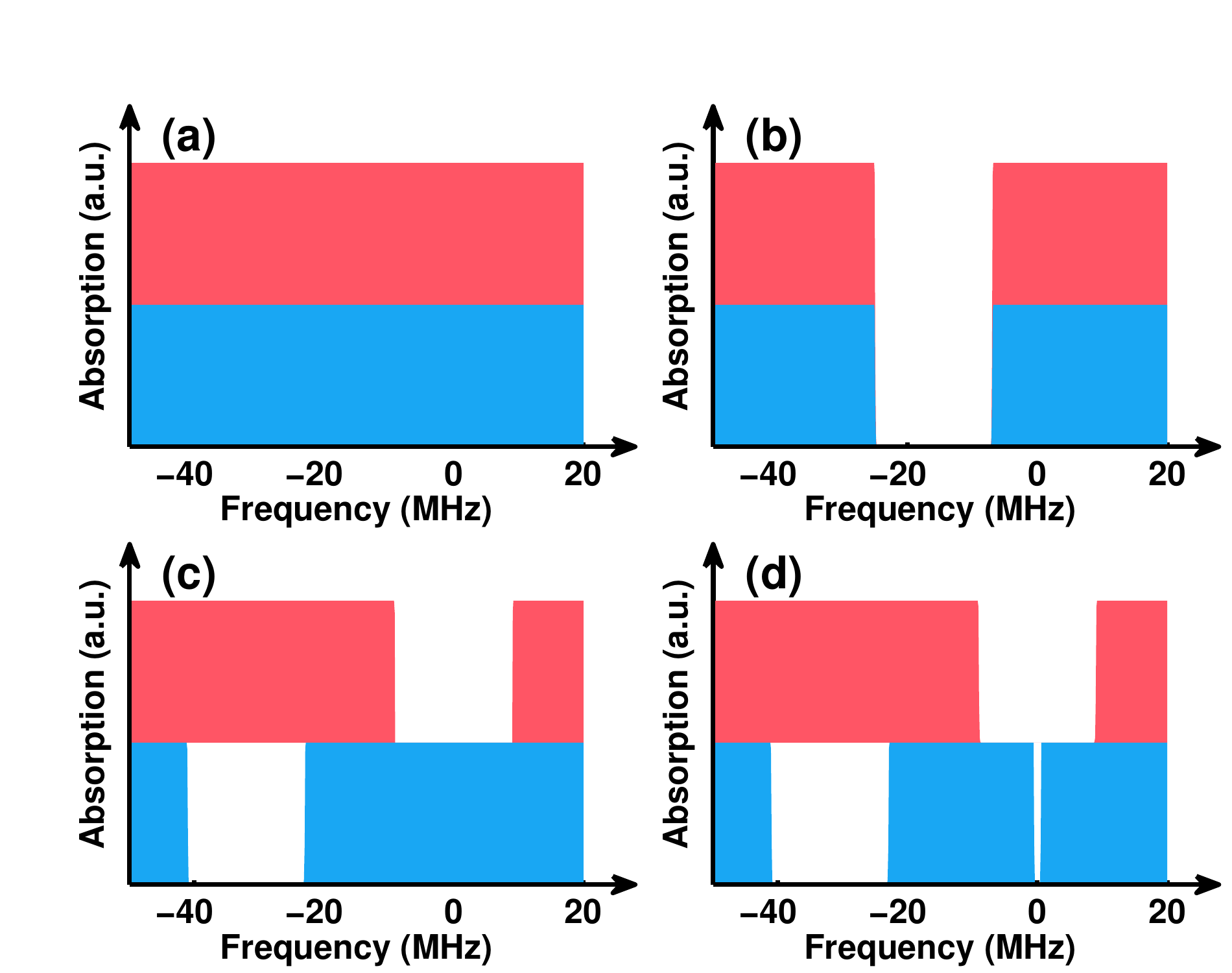}
\caption{ Absorption profile of the ions during the preparation of the frequency shift filter. The blue/red color represents absorption from ions with a positive/negative sign of their Stark coefficient (a) The original absorption profile when V = 88 V. (b) A hole of about 18 MHz was prepared by spectral hole burning when V = 88 V. (c) The 18 MHz hole is split in two when V = 0. (d) Another 1 MHz narrow hole is burnt at around 0 frequency. The 1 MHz hole which only surrounded by ions with the same sign of the Stark coefficient within the 18 MHz region is called the frequency shift filter (frequency shifter). }
\label{fig:filter_all}
\end{figure}

The sketch of a step by step absorption profile is shown in Fig. \ref{fig:filter_all}, where the red color represents the ions with a negative sign of their Stark coefficient while the blue color represents the ions with a positive sign of their Stark coefficient relative to the E field. An external voltage of 88 V was applied prior to any preparation, shifting the blue ions 16 MHz and the red ones $-16$ MHz, the absorption profile of the interested frequency range is shown in Fig. \ref{fig:filter_all} (a). Then an 18 MHz wide hole was prepared using the optical pumping scheme described in our previous paper \cite{S_RefA, S_RefB}. The resulting absorption profile is shown in Fig. \ref{fig:filter_all} (b). The external field was switched off, the ions will shift back to their original position, left two 'half holes' at 0 MHz and $-32$ MHz, shown in Fig. \ref{fig:filter_all} (c), so that at the region of $0\pm9$ MHz only the ions with a positive sign of their Stark shift exist. Lastly, the laser frequency was scanned $\pm500$ kHz around 0 MHz. In the end, a $\sim$1 MHz narrow hole was created with the surrounding ions within 18 MHz having the same sign of their Stark coefficient was created. This is the frequency shifter (frequency shift filter) and it is shown as the structure around 0 MHz in Fig. \ref{fig:filter_all} (d). When an external voltage applied, the frequency shift of the $\sim$1 MHz hole will be proportional to the E field. 

 \subsection{E field}

\begin{figure}[htbp]
\centering
\includegraphics[width=\linewidth]{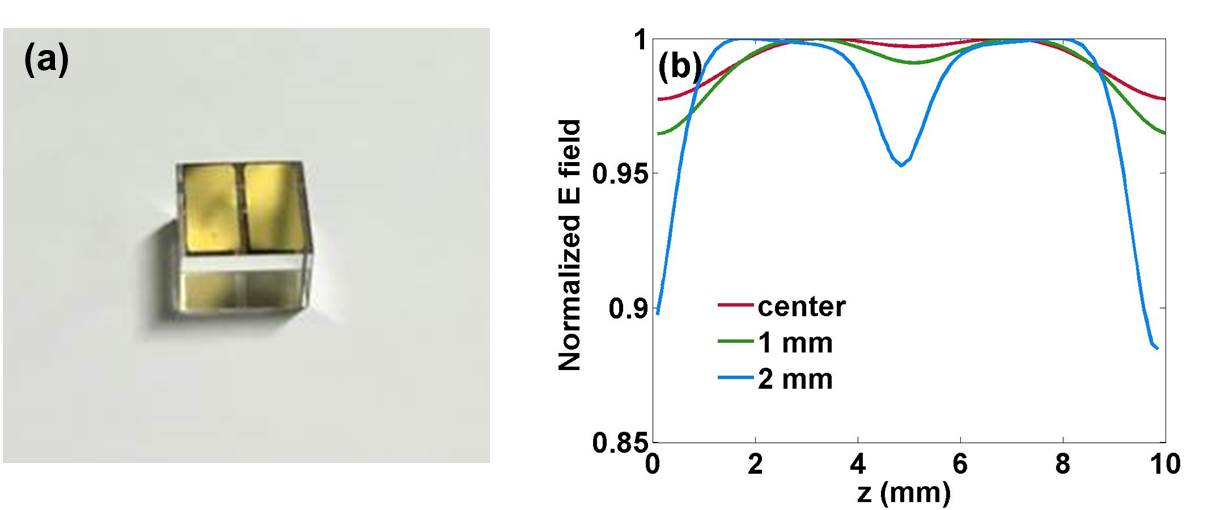}
\caption{(a) The 6 mm thick crystal with electrodes used in the experiment. Two sets of electrodes were directly deposited onto the top and bottom surfaces of the crystal with a 0.5 mm gap in between them. (b) The electric field along the light propagation axis at the center of the crystal (red), 1 mm away (green) and 2 mm away (blue) from the center. As can be seen there is a spatial field inhomogeneity which is especially pronouns when close to the electrode surfaces. }
\label{fig:crystal_Efield}
\end{figure}

Two sets of electrodes were directly deposited onto the top and bottom surfaces of the crystal with a 0.5 mm gap in between the two sets of electrodes as shown in Fig. \ref{fig:crystal_Efield} (a). Because of the two pairs of electrodes separated by a gap on each of the two surfaces, the E field in the crystal will be nonuniform when an external voltage is applied across the crystal. Hence the amount of Stark shift will be slightly different for ions in different spatial positions. The E field along the light propagation path is shown in  Fig. \ref{fig:crystal_Efield} (b) about $0.2\%$ inhomogeneity is shown in the center of the crystal along the light propagation path, it becomes more nonuniform when away from the center. The light propagation was close to the center in the experiment to get a better uniformity of the E field.

\end{document}